\documentclass[twocolumn,showpacs,preprintnumbers,prb,fleqn,superscriptaddress]{revtex4-2}
\usepackage[english]{babel}
\usepackage[utf8]{inputenc}
\usepackage{SIunits}
\usepackage{graphicx}
\usepackage{dcolumn}
\usepackage{amsmath}
\usepackage{natbib}
\usepackage{bm}
\usepackage{textcomp}
\usepackage{color}
\usepackage{float}
\usepackage{booktabs}
\usepackage{multirow}
\usepackage{tabularx}
\usepackage{adjustbox}
\usepackage{hyperref}
\usepackage{tikz}
\usepackage{makecell}

\begin{document}

\title{Raman scattering signatures of the strong spin-phonon coupling in the bulk magnetic van der Waals material CrSBr}

\author{Amit Pawbake}
\email{amit.pawbake@lncmi.cnrs.fr}\affiliation{LNCMI, UPR 3228, CNRS, EMFL, Université Grenoble Alpes, 38000 Grenoble, France}
\author{Thomas Pelini}
\affiliation{LNCMI, UPR 3228, CNRS, EMFL, Université Grenoble Alpes, 38000 Grenoble, France}
\author{Nathan P. Wilson}
\affiliation{Walter Schottky Institut, Physics Department and MCQST, Technische Universitat Munchen, 85748 Garching, Germany}
\author{Kseniia Mosina}
\affiliation{Chemistry Department, University of Chemistry and Technology Prague, 16628 Prague,
Czech Republic}
\author{Zdenek Sofer}
\affiliation{Chemistry Department, University of Chemistry and Technology Prague, 16628 Prague,
Czech Republic}
\author{Rolf Heid}
\affiliation{Institute for Quantum Materials and Technologies, Karlsruhe Institute of Technology, 76131 Karlsruhe, Germany}
\author{Clement Faugeras}
\email{clement.faugeras@lncmi.cnrs.fr} \affiliation{LNCMI, UPR 3228, CNRS, EMFL, Université Grenoble Alpes, 38000 Grenoble, France}

\date{\today }

\begin{abstract}

Magnetic excitations in layered magnetic materials that can be thinned down the two-dimensional (2D) monolayer limit are of high interest from a fundamental point of view and for applications perspectives. Raman scattering has played a crucial role in exploring the properties of magnetic layered materials and, even-though it is essentially a probe of lattice vibrations, it can reflect magnetic ordering in solids through the spin-phonon interaction or through the observation of magnon excitations. In bulk CrSBr, a layered A type antiferromagnet (AF), we show that the magnetic ordering can be directly observed in the temperature dependence of the Raman scattering response i) through the variations of the scattered intensities, ii) through the activation of new phonon lines reflecting the change of symmetry with the appearance of the additional magnetic periodicity, and iii) through the observation, below the N\'{e}el temperature ($T_N$) of second order Raman scattering processes. We additionally show that the three different magnetic phases encountered in CrSBr, including the recently identified low temperature phase, have a particular Raman scattering signature. This work demonstrates that magnetic ordering can be observed directly in the Raman scattering response of bulk CrSBr with in-plane magnetization, and that it can provide a unique insight into the magnetic phases encountered in magnetic layered materials.
\end{abstract}

\pacs{73.22.Lp, 63.20.Kd, 78.30.Na, 78.67.-n}

\maketitle

\section{Introduction}

Layered materials that can be thinned down to the monolayer limit and presenting magnetic properties, have emerged as platforms to engineer and investigate exotic magnetic ground states in low dimensions~\cite{Wang2022}. The observation of long range magnetic order in two dimensional monolayers of CrI$_3$~\cite{Huang2017} and of Cr$_2$Ge$_2$Te$_6$~\cite{Gong2017} have triggered a large number of investigations aiming at finding other layered magnetic 2D materials, and at using them within van der Waals (vdW) heterostructures to introduce magnetic properties through proximity effects~\cite{Zhong2017,Ciorciaro2020}, and for spintronics applications such as spin valves/filters. They are also foreseen as potential building blocks for composite Qubits for quantum technologies~\cite{Li2020}. These layered magnetic materials present a broad variety of magnetic orders with intra and inter-layer ferro- or antiferromagnetic interactions. Taking into account the variety of possible antiferromagnetic states (N\'{e}el-like, zig-zag, stripe) there is a need to establish experimental techniques sensitive to these different orders and symmetries. Within this broad family of materials, CrSBr has recently stimulated a large interest because this material can be thinned down to the monolayer with persisting magnetic properties and, in contrast to CrX$_3$ or VX$_3$ compounds (where X=I, Br or Cl), bulk CrSBr is stable in air and can hence be manipulated in standard conditions~\cite{Telford2020}.

The crystal structure of bulk CrSBr is orthorhombic (see Fig~\ref{Fig1}) with space group \textit{Pmmn (D$_{2h}$)} and no structural phase transitions between $15$~K and $300$~K~\cite{Telford2020}. From optoelectronics perspectives, it is a direct band gap semiconductor with E$_g=1.5$~eV and hosts tightly bound excitons that give rise to photoluminescence signals close to $1.3$~eV~\cite{Telford2020,Wilson2021}. The electronic properties of CrSBr are strongly anisotropic and this is reflected in its optical~\cite{Wilson2021,Klein2022} and transport properties~\cite{Wu2022}. From these different viewpoints, this material can be seen as a quasi-1D system. CrSBr is a layered materials with strong ferromagnetic intralayer interactions which align the Cr$^{3+}$ spins within the plane of the layers, along the crystallographic b-axis (easy axis), see Fig.~\ref{Fig1}. This intralayer ferromagnetic order appears below $T_C=160$~K for bulk samples~\cite{Lee2021}. The Cr$^{3+}$ ions couple antiferromagnetically across the van der Waals gap and bulk CrSBr is an A type antiferromagnet with $T_N=133$~K~\cite{Gser1990}, as evidenced by magnetization and magnetotransport~\cite{Telford2020} investigations. In between $T_C$ and $T_N$, bulk CrSBr is in an intermediate magnetic phase (iFM) in which the intralayer ferromagnetic order is well established while the interlayer ordering is only partial, comprising both ferro- and antiferromagnetic interlayer couplings~\cite{Lee2021,Liu2022}. When lowering the temperature below $T^*=40$~K, a change of sign of the magnetoresistance with a simultaneous increase of the magnetization~\cite{Telford2021}, and muon spin relaxation experiments~\cite{LopezPaz2022} have pointed towards a new magnetic phase. The details of this phase are still under debate and it has been attributed either to the ferromagnetic ordering of magnetic defects within the bulk crystal~\cite{Telford2021,Klein2023} or, alternatively, to a gradual decrease of spin fluctuations below $T=100$~K leading to a spin freezing process at $T^*=40$~K with potentially the formation of domains with different magnetic ground states~\cite{LopezPaz2022}.

Antiferromagnetic magnon excitations~\cite{Rezende2019} in bulk CrSBr and their dispersion when applying external magnetic fields have been measured recently with electron paramagnetic resonance (EPR) spectroscopy~\cite{Cham2022} or through their coupling to excitons in time resolved reflectivity spectroscopy~\cite{Bae2022}. The magnon spectrum at $B=0$~T is composed of two excitations at $25$~GHz and $35$~GHz reflecting the biaxial magnetic anistotropy of CrSBr. If the direct magnon absorption measured in EPR is well understood, the coupling of magnons to excitons through the strain field induced by the heating due to a pump pulse implies the magneto-elastic interaction and phonons of bulk CrSBr. These measurements provide qualitatively similar behaviors but different energies for magnons~\cite{Bae2022} than those extracted from direct absorption~\cite{Cham2022} leading, \textit{in-fine}, to significantly different microscopic magnetic parameters. A detail understanding of this interaction hence appears essential to describe the magnetic properties of materials with in-plane magnetization, for which standard Kerr effect measurements, mostly sensitive to the $M_z$ component of the magnetization, can hardly be applied and alternative approaches have to be implemented.

In this article, we use Raman scattering spectroscopy to detect the magnetic phase transitions in bulk CrSBr when lowering temperature. By tilting the sample with respect to the excitation light propagation direction, we activate Raman scattering from phonons at the Z point of the Brillouin zone which are particulary sensitive to the AF ordering along the c-axis. The strong spin-lattice coupling is evidenced through the appearance of new phonon modes when the additional magnetic periodicity develops below $T_N$ and by pronounced variations of the scattered intensity. These changes in the Raman scattering spectrum allow identifying the intermediate magnetic phase when $T_C > T > T_N$, the AF order, and also the hidden order phase below $T^*$. We show that the evolution of phonon modes when lowering temperature from room temperature down to liquid helium temperature cannot be described by usual lattice dynamics~\cite{Zhu2021} and strongly reflects the magnetic ordering and the different magnetic phases in this bulk compound. Additionally, we apply an external magnetic field which effectively modifies the Raman scattering response, further demonstrating the spin-phonon coupling in bulk CrSBr.

\section{Methods}

\subsection{Samples and experimental setup}

The CrSBr single crystals were synthesized through a chemical vapour transport (CVT) method. Chromium, sulfur and bromine in a stoichiometric ratio of CrSBr were added and sealed in a quartz tube under a high vacuum ($50$~g of CrBrS with $0.1\%$ Br excess, ampoule size $50x250$~mm). The tube was, then, placed into a two-zone tube furnace. The pre-reaction was done in crucible furnace where one end of ampoule was keept below $300^{\circ}$C and bottom was gradually heated on $700^{\circ}$C over a period of $50$~hours. Then, the source and growth ends were kept at $800$ and $900^{\circ}$C, respectively. After 25 hours, the temperature gradient was reversed, and the hot end gradually increased from $850$ to $920^{\circ}$C over 10 days. High-quality CrSBr single crystals with lengths up to $2$~cm were achieved.
Bulk CrSBr was placed on the cold finger of a variable temperature helium flow cryostat. We use a long working distance optical objective with numerical aperture of $0.55$ to excite the sample and collect Raman scattering signals.
Raman scattering signals were analyzed by a grating spectrometer equipped with a nitrogene cooled silicon Charge Coupled Device (CCD). We use a depolarizer to excite with unpolarized excitation and to detect the unpolarized Raman scattering response. Measurements were performed at liquid helium temperature with a $\lambda=633$~nm excitation from solid state laser diodes, keeping the optical power below $0.3$~mW. For magneto-Raman or magneto-photoluminescence measurements we have used a home-made experimental set-up based on free beam propagation of optical excitation and collection. We use a long working distance objective with a numerical aperture $NA=0.35$ used to focus the excitation beam down to a spot size of $1$~$\mu$m and to collect Raman scattering or photoluminescence signals. The sample was placed on piezo motors allowing for the spatial mapping of the optical response. This set-up is then inserted in a close metallic tube fuilled with helium exchange gas and then placed at liquid helium temperature in a superconducting solenoid producing magnetic fields up to $B=14$~T.

\subsection{Theory}
Lattice dynamics properties for orthorhombic CrSBr were calculated using the linear response or density-functional perturbation theory (DFPT) implemented in the mixed-basis pseudopotential method~\cite{Meyer,Heid1999}. The electron-ion interaction is described by norm-conserving pseudopotentials, which were constructed following the descriptions of Vanderbilt~\cite{Vanderbilt1985}.  Semi-core states Cr-3s, Cr-3p were included in the valence space. In the mixed-basis approach, valence states are expanded in a combination of plane waves and local functions at atomic sites, which allows an efficient description of more localized components of the valence states. Here, plane waves with a cut-off for the kinetic energy of $24$~Ry and local functions of $s$,$p$ type for S and Br, and $s$, $p$, $d$ type for Cr, respectively, were employed. Spin-polarized calculations were done for both the small cell assuming a ferromagnetic (FM) ground state, and a $1\times 1\times 2$ supercell with an antiferromagnetc (AFM) ground state. Brillouin-zone integration was performed by sampling a $12\times 8\times 4$ $k$-point a mesh for the FM cell and a $12\times 8\times 2$ (AFM) $k$-point mesh for the AFM cell, respectively. The exchange-correlation functional was represented by the general-gradient approximation (GGA) in the PBE form~\cite{Perdew1996}.

Within GGA, the ground state of CrSBr in the magnetic state is insulating, albeit with a rather small gap size of $0.3$~eV, as compared to the experimental optical gap of $1.25$~eV~\cite{Telford2020}. To improve on the gap size, correlations at the Cr site were incorporated in the DFT+U scheme. Following previous publications~\cite{Wang2020,Yang2021,Xu2022}, we used $U=4$~eV and $J=1$~eV, (corresponding to $U_{eff}=U-J$ of $3$ eV), thereby increasing the gap to $1.3$~eV. The main effect of the +U correction on the phonons was a hardening in particular of the high-frequency modes due to an reduced electronic screening.

Lattice constants for the orthorhombic structure were taken from measurements at $T=10$~K by Lopez-Paz et al. ($a=3.51069$~\AA, $b=4.74623$~\AA, $c=7.9162$~\AA)~\cite{LopezPaz2022}. Internal structural parameters were relaxed until the atomic forces
were smaller than $2.6\times 10^{-2}$~eV/\AA. Above mentioned calculational parameters guaranteed convergence of phonon frequencies better than 1$cm^{-1}$. Comparison of phonons for the FM and AFM cell showed that the difference in magnetic structure has hardly any effect on phonon frequencies. The main difference is a folding of the $Z$-point phonons of the FM cell to the $\Gamma$ point of the AFM phase. This renders some of the cell modes at $Z$ (FM) to become Raman active in the AFM phase. Full phonon dispersion for the FM cell was calculated by performing DFPT calculations of dynamical matrices on a $4\times 4\times 2$ momentum mesh, and subsequently using standard Fourier interpolation to obtain dynamical matrices throughout the Brillouin zone. To correctly describe the LO/TO splitting of infrared active modes, the nonanalytic part of the dynamical matrices was included following the procedure outlined in Ref.~\cite{Sklyadneva2012}. It does not effect the Raman modes.

\section{Results and discussions}

\begin{figure}
\includegraphics[width=1\linewidth,angle=0,clip]{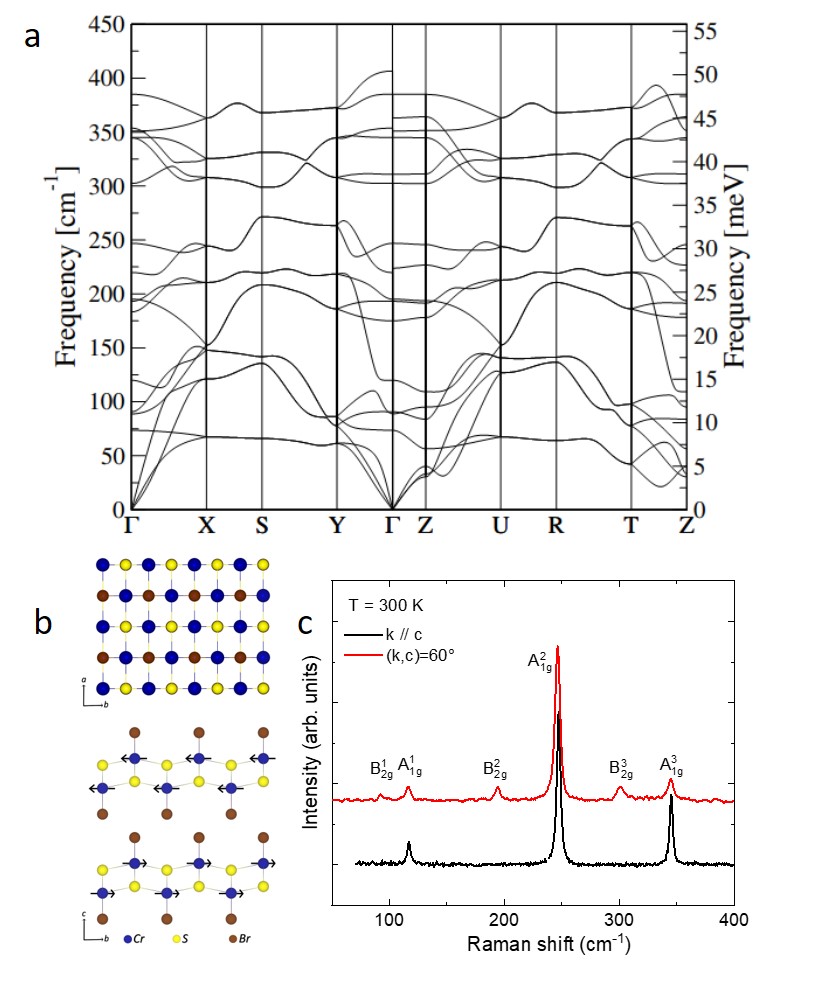}
\caption{a) Calculated phonon band structure of bulk CrSBr at $T=10$~K. b) Crystal structure of CrSBr as seen along the c or a crystallographic axis. The arrows represent the magnetic moments aligned in the plane of the layers. c) Room temperature Raman spectra of bulk CrSBr in perpendicular configuration (black curve) or with a tilt angle of $60^{\circ}$ between the light propagation direction and the normal to the surface of the crystal (red curve).
\label{Fig1}}
\end{figure}

The primitive cell of bulk CrSBr includes $6$ atoms in the high temperature paramagnetic phase, and $12$ atoms in the AF super cell (1x1x2) below T$_N$. The phonon spectrum includes 15 optical modes and 3 acoustical branches. The calculated phonon Brillouin zone for bulk CrSBr is presented in Fig.~\ref{Fig1}a. When measured in back scattering geometry with the incident light perpendicular to the surface, the unpolarized first order Raman scattering spectrum of bulk CrSBr is composed of three main contributions at $117$, $257$, and $345$~cm$^{-1}$, respectively (see Fig~\ref{Fig1}c). They arise from $\Gamma$-point optical phonons of A$_{1g}$ symmetry, with atomic displacements pattern perpendicular to the plane of the layers. When imposing a $60^{\circ}$ tilt angle to the sample, three new contributions appear in the room temperature Raman scattering response, at $92$, $194$ and $301$ cm$^{-1}$ which correspond to the three B$_{2g}$ modes at the $\Gamma$ point, with atomic displacements in the plane of the layers.

\begin{figure}
\includegraphics[width=1\linewidth,angle=0,clip]{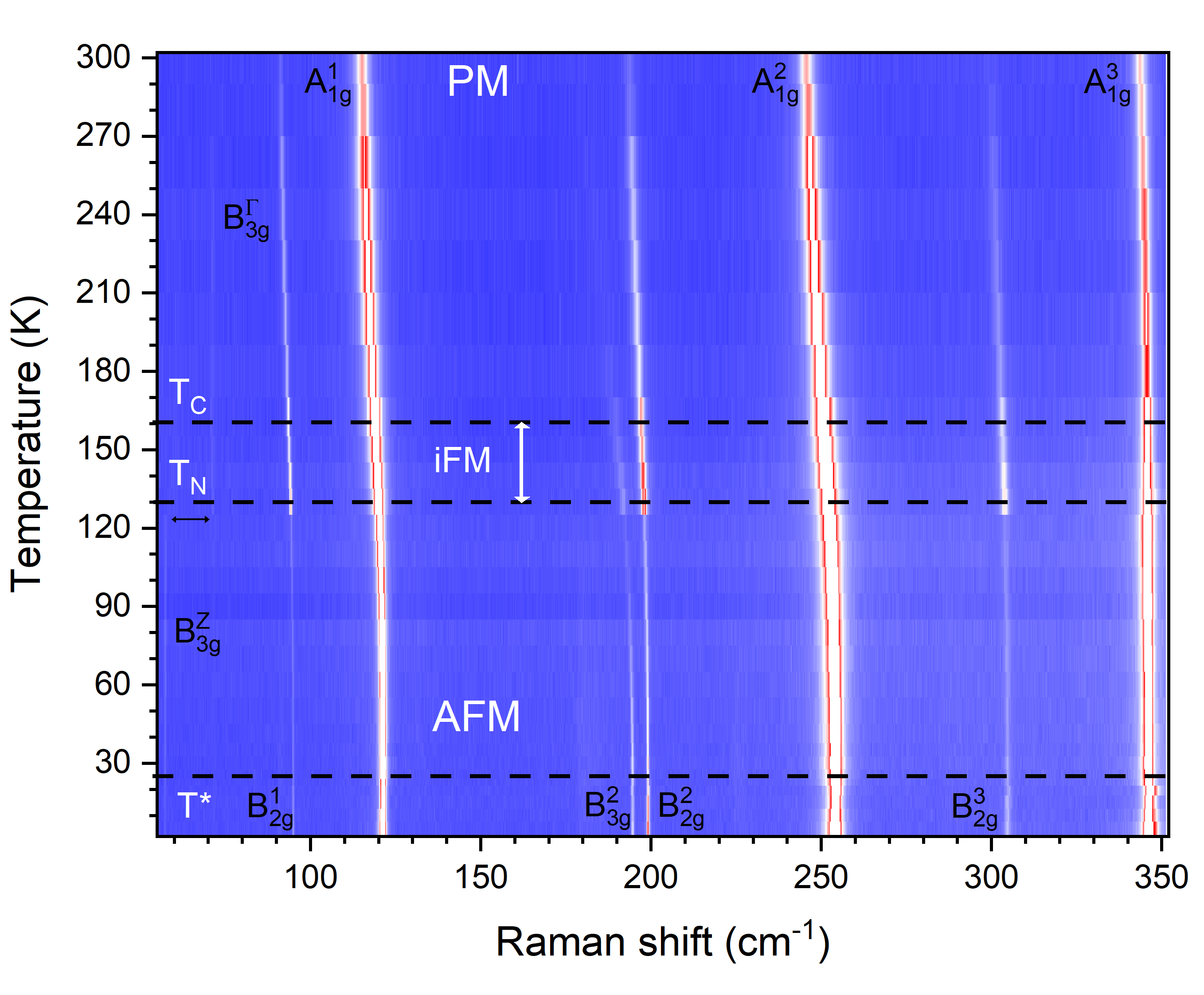}
\caption{False color map of the unpolarized Raman scattering response of bulk CrSBr as a function of temperature.  The different magnetic phases are indicated, the paramagnetic phase (PM), the intermediate magnetic phase (iFM) and the antiferromagnetic phase (AF), together with the critical temperatures T$_N$ and T$^*$ and the different phonon symmetries.
\label{Fig2}}
\end{figure}

\begin{figure}
\includegraphics[width=1\linewidth,angle=0,clip]{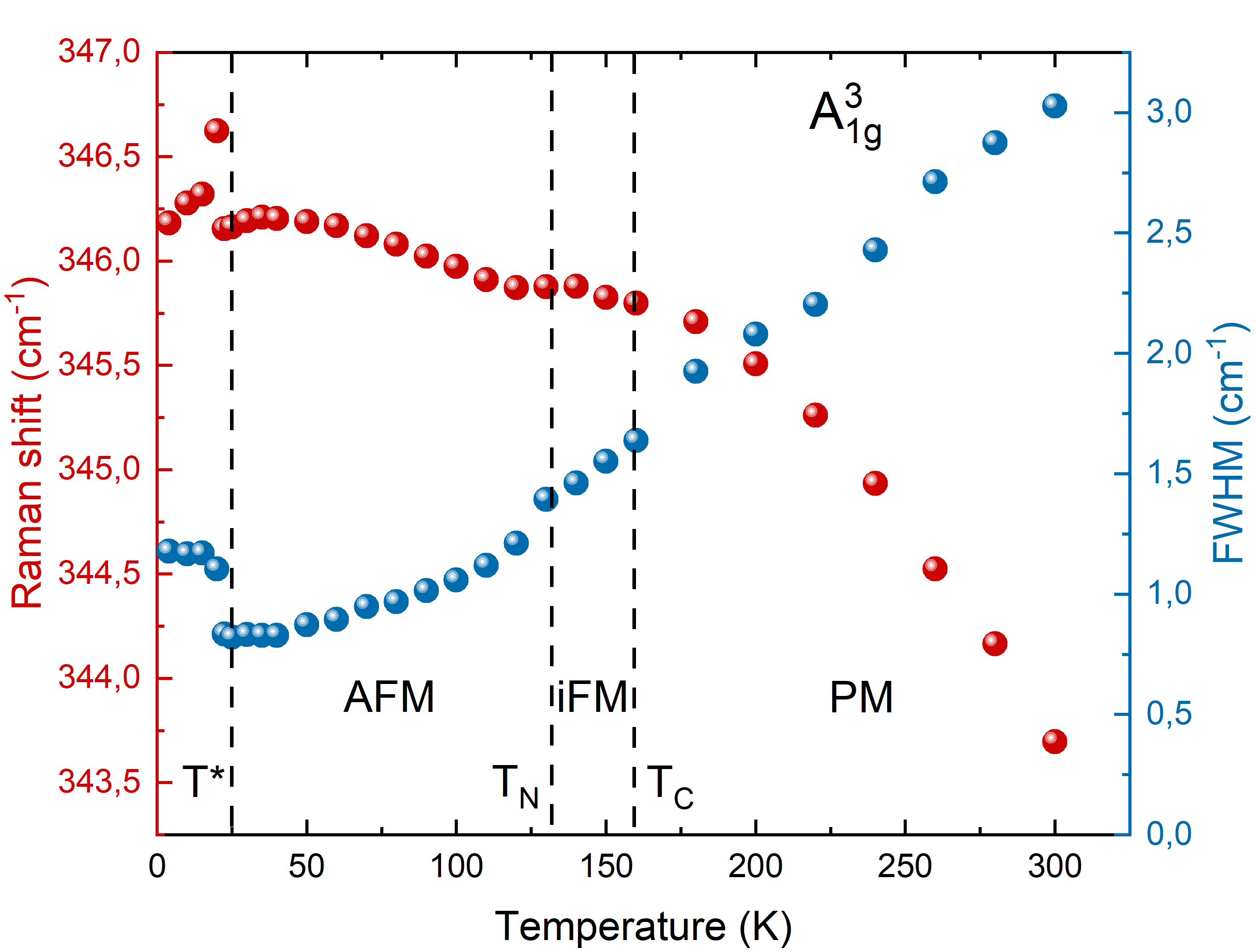}
\caption{Evolution of the $A_{1g}^3$ phonon energy (red points) and linewidth (blue points) as a function of temperature. The errors are smaller than the symbol size.
\label{Fig3}}
\end{figure}

The experimentally observed evolution of phonon energies in CrSBr as a function of temperature, presented in Fig.~\ref{Fig2}, clearly goes beyond phonon evolutions observed in semiconducting 2D materials~\cite{Zhu2021}. When lowering temperature, anharmonicity effects lead to a decrease of the phonons linewidths and to an increase of the phonons energies~\cite{Balkanski1983,Kolesov2016} with a dependence that reflects multiphonon processes and that, in a first approximation, takes the form $\omega(T)=\omega_0+c_i <n_i>$, where $<n_i>$ is the temperature dependent population of phonons with energy $\omega_i$ and c$_i$ is a numerical factor~\cite{Kolesov2016}. At $T=5$~K, the three main Raman scattering features corresponding to the $A_{1g}^{1,2,3}$ phonons are observed at $121$, $254$ and $346$ cm$^{-1}$, respectively. A comparison between experimental and calculated phonon energies is presented in Table~$S1$ of the Supplementary Materials. Let us first investigate the evolution of the $A_{1g}^3$ phonon with temperature presented in Fig.~\ref{Fig3}. Below $T_N$, CrSBr becomes magnetically ordered and the spin-spin interaction is modified by propagating phonons that can change the superexchange angles and the resulting the magnetic exchange couplings~\cite{Vaclavkova2020}. This interaction might also slightly modify the lattice constant and phonon energies~\cite{Xu2022}. We observe a hardening of the phonon mode from room temperature down to $T_C$ where the phonon energy stays rather constant down to $T_N$. This temperature range corresponds to the intermediate magnetic phase in which the intralayer ferromagnetism is established, but spins in adjacent layers are not all AF ordered~\cite{Lee2021,Liu2022}. For lower temperatures, the $A_{1g}^3$ phonon mode hardens again before showing a sharp discontinuity around $T=25$~K which coincides with the transition to a new magnetic ground state~\cite{Telford2021,LopezPaz2022}. The evolution of the Full Width at Half Maximum (FWHM) of the A$_{1g}^3$ phonon, presented in Fig.~\ref{Fig3} also deviates from expectations considering only anharmonicity: The FWHM decreases when lowering temperature but shows an inflection point in the intermediate magnetic phase, around $T=140$~K. We understand this behavior as a combined effect of anharmonicity with the reduction of the phonon thermal populations, and of the simultaneous reduction of spin fluctuations of the magnetic ions~\cite{LopezPaz2022} which leads to the suppression of a phonon scattering mechanism and to a decrease of the linewidth. Close to $T=25$~K, we observe a sharp increase of the FWHM of the $A_{1g}^3$ phonon. The evolution of both the energy and of the FWHM of the A$_{1g}^3$ phonon hence reflects the details of the magnetic ordering in this bulk magnetic van der Waals material, revealing the four magnetic phases discovered in bulk CrSBr.

In addition to these effects, particularly pronounced for the $A_{1g}^3$ phonon mode\footnote{This behavior is also observed in a less pronounced way with phonons $B_{2g}^2$, $A_{1g}^2$ and $B_{2g}^3$, see Fig.~S1-S2 of Supplementary Materials.}, the emergence of magnetism in bulk CrSBr and the spin-phonon coupling also has profound implications on the overall Raman scattering response. As can be seen in Fig.~\ref{Fig2}, the scattered intensity from the different phonon modes increases significantly for $130~K < T < 160~K$ which corresponds to the intermediate magnetic phase. We present in the Supplementary Materials the evolution of the energies, width and integrated intensities of all observed Raman scattering features, see Fig.~S1-S3. The evolution of the phonon energies goes beyond effects of crystal anharmonicity and most of the peaks show intensity anomalies corresponding to the iFM phase when $T_N < T < T_C$. The increase of fullwidth below $T=25$~K, particularly marked on the $A_{1g}^3$ phonon, is also observable on $A_{1g}^2$, but not on the $A_{1g}^1$ phonon. The increase of the scattered intensity in the iFM phase affects most of the phonon modes (see Supplemental Material Fig.~S3).

\begin{figure}
\includegraphics[width=1\linewidth,angle=0,clip]{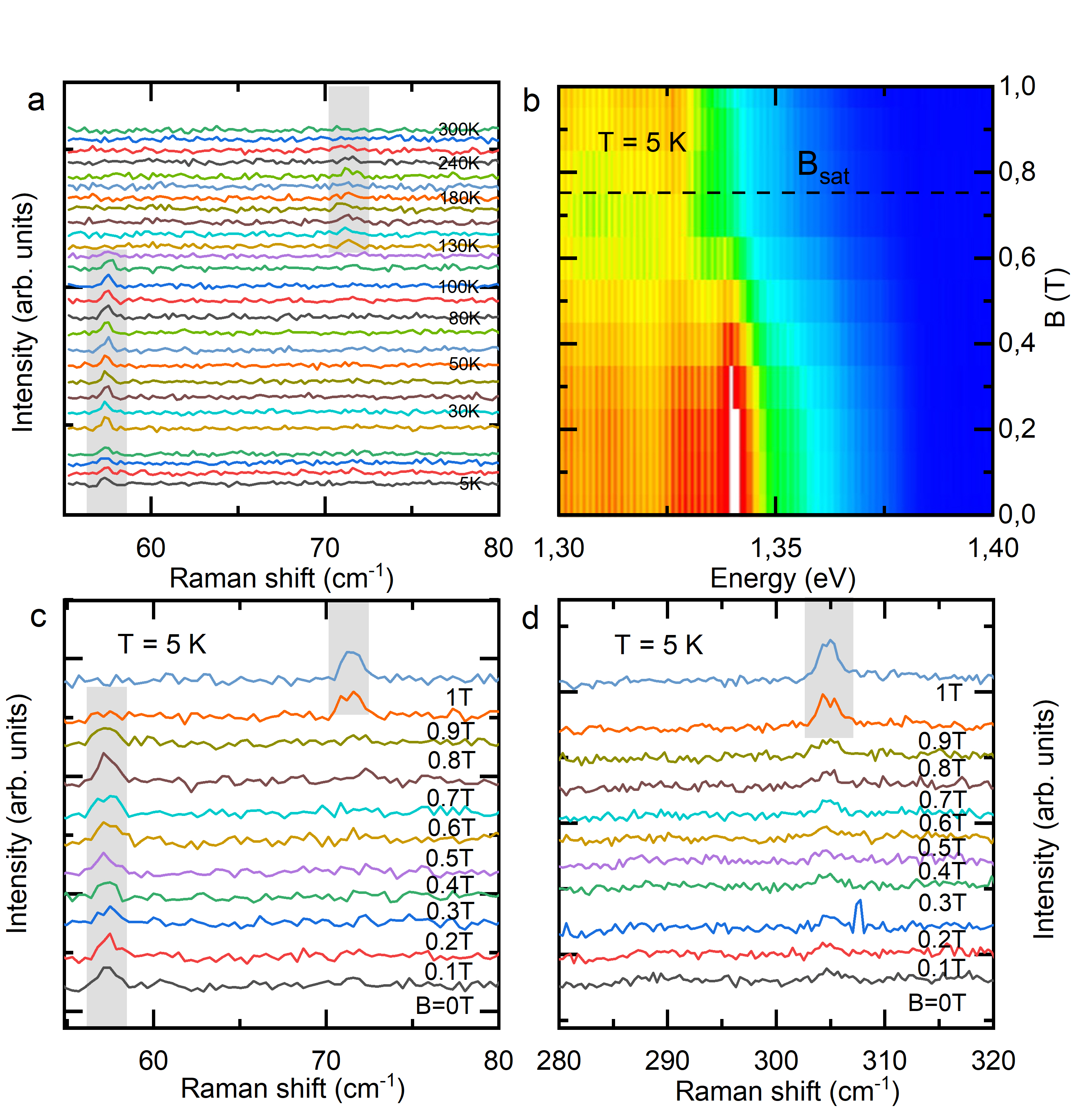}
\caption{a) Evolution of the unpolarized Raman scattering response measured at various temperatures between $T=5$~K up to $T=300$~K, b) Magneto-photoluminescence of bulk CrSBr up to $B=1$~T with the B field applied $30^{\circ}$ from the c axis and showing a saturation field of $B_{sat}=0.75$~T, c,d) Magneto-Raman scattering response at $T=5$~K with the B field applied in the same configuration as in panel b) for two different energy windows. The shaded regions indicate the energy position of the  B$_{3g}$ phonon modes.
\label{Fig4}}
\end{figure}

When the interlayer AF order builds up, the symmetry of the solid changes due to the additional magnetic periodicity imposed on the crystallographic one. In bulk CrSBr, the unit cell doubles along the c axis which causes the folding of Z-point phonons on the $\Gamma$-point and the activation of new phonon modes in the Raman scattering response. This effect can be clearly seen with the gradual appearance, for temperatures close to $T_C$, of an additional phonon peak close to $194$~cm$^{-1}$ which we identify as the  B$_{3g}$ phonon from the Z point. Additionally, a weak Raman scattering mode at $71$~cm$^{-1}$ is observed below $T=260$~K and corresponds to a B$_{3g}$ phonon from the $\Gamma$ point. At $T=130$~K, this mode disappears and a new mode rises at $57$~cm$^{-1}$, see Fig.~\ref{Fig4}a. We identify this new mode as the B$_{3g}$ mode from the Z point, activated below T$_N$. To further reenforce our interpretation that the appearance this phonon below T$_N$ is related to the AF order, we have applied a magnetic field at low temperature. The sample in its AF phase and the c-axis is tilted by $30^{\circ}$ with respect to the light propagation direction and to the external magnetic field. As the bulk sample is not oriented in the (a,b) plane and we have simultaneously measured the photoluminescence from excitons in bulk CrSBr, which have been shown to be a probe of the saturation field~\cite{Wilson2021} above which all the spins are aligned along the direction of the external magnetic field. These results are shown in Fig.~\ref{Fig4}b,c,d) for the unpolarized magneto-photoluminescence and magneto-Raman scattering, respectively. When measuring the magneto-Raman scattering response, we observe that the $57$~cm$^{-1}$ mode observed in the AF phase gradually disappears when the saturation field is reached, in our tilted configuration for $B>0.8$~T, and the $71$~cm$^{-1}$ phonon mode is recovered. Simultaneously, the $B^2_{2g}$ phonon observed at $305$~cm$^{-1}$ strongly gains in intensity for $B>0.8$~T. This magnetic field corresponds to the saturation magnetic field extracted from magneto-photoluminescence and presented in Fig.~\ref{Fig4}b, confirming that observing the $B_{3g}$ from Z point at $57$~cm$^{-1}$ is a characteristic signature of the AF phase. The results presented in Fig.~\ref{Fig4}c and d hence clearly relate the evolution of the Raman scattering response to the magnetic ordering in bulk CrSBr. Puzzling is the complete disappearance of the B$_{3g}$ phonon from the $\Gamma$ point below $T_N$ accompanied by the strong reduction of intensity of the B$_{2g}$ phonon from the $\Gamma$ point at $290$~cm$^{-1}$. In the case of phonon zone folding due to the additional magnetic periodicity below $T_N$, one would expect the appearance of new phonon modes in the Raman scattering response of the magnetic phase, but phonons already active in the high temperature phase are expected to persist which is not the case in bulk CrSBr pointing towards a more complex process.

\begin{figure}
\includegraphics[width=1\linewidth,angle=0,clip]{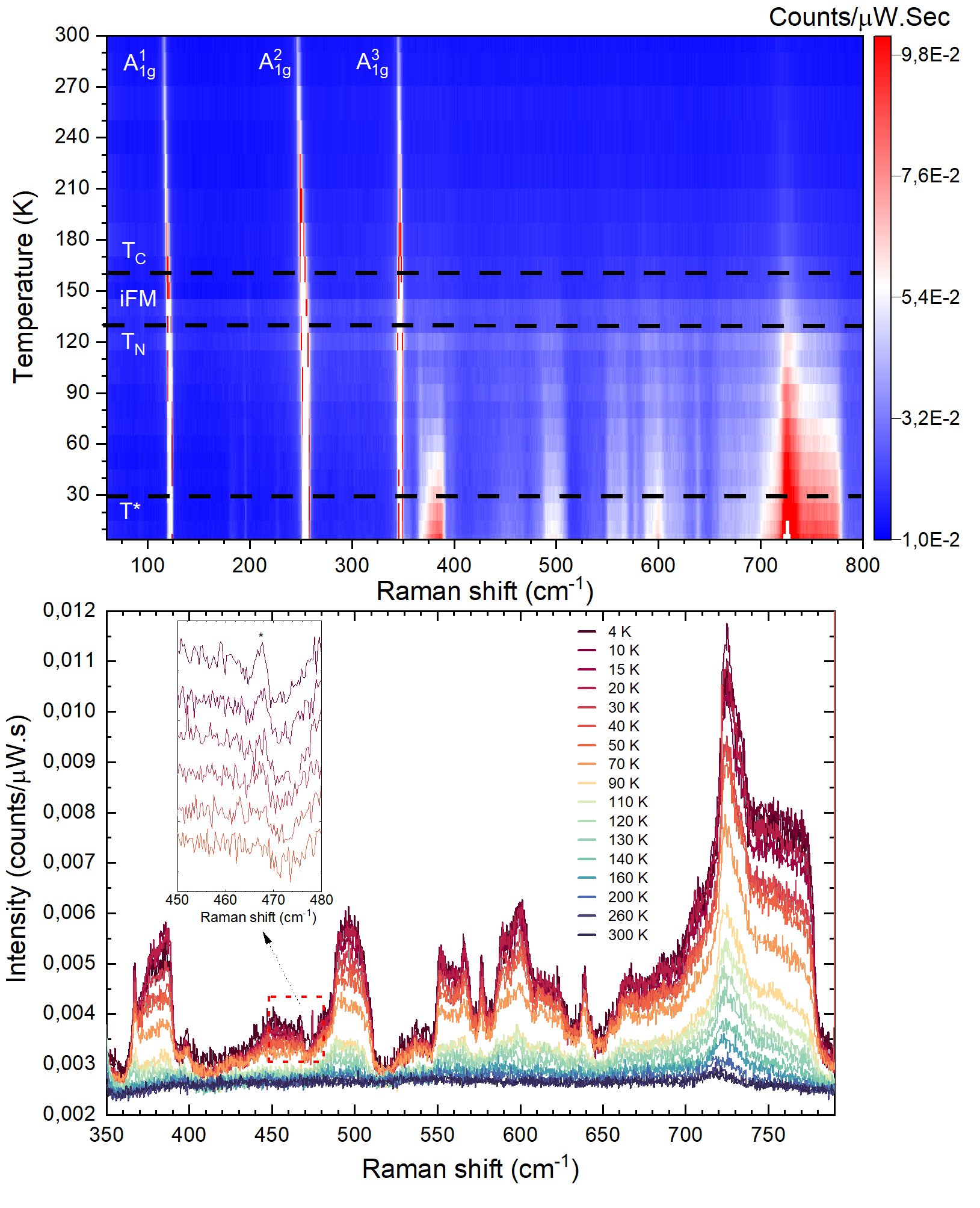}
\caption{Top: False color map of the Raman scattering response of bulk CrSBr as a function of temperature showing both the first and second order Raman scattering spectrum. Bottom: Raman scattering spectra at different temperatures above and below T$_N$. The inset focuses on the peak at $467$~cm$^{-1}$ (indicated by a star) that appears below T$^*$.
\label{Fig5}}
\end{figure}

When the magnetic order appears in bulk CrSBr, the first order Raman scattering spectrum is profoundly changed as described above, but we also observe the appearance of the second order Raman scattering spectrum. This evolution is presented in Fig.~\ref{Fig5} a and b. The second order appears when the temperature is tuned below $160$~K which corresponds to the establishment of the intralayer ferromagnetic order, and grows in intensity when temperature is further decreased. It extends from $370$~cm$^{-1}$ to $780$~cm$^{-1}$ and includes several broad features, with width $20-30$~cm$^{-1}$ which correspond to multiphonon processes, including acoustical phonons at the Z and X points of the phonon Brillouin zone (see Fig.~\ref{Fig1}a). The appearance of the second order spectrum when decreasing temperature is related to a growing resonance of the laser energy, in this case $E_{exc}=1.96$~eV, with the B-exciton energy at $1.82$~eV.

We have previously described the sharp increase of the energy and linewidth of the $A_{1g}^3$ phonon below $T=25$~K. The low temperature magnetic phase has another Raman scattering signature within this second order spectrum. As we show in the inset of Fig.~\ref{Fig5}b, we observe at $E=467$~cm$^{-1}$ an additional sharp Raman scattering peak with FWHM$=2$~cm$^{-1}$, that can only be observed at the lowest temperatures, when $T < 40$~K. This energy has an energy higher than the first order phonon Raman scattering spectrum of bulk CrSBr, and can hence hardly be assigned to the activation of a previously forbidden phonon Raman scattering and would rather be a multi-phonon process involving for instance the simultaneous emission of a $A_{1g}^{1}$ and of a $A_{1g}^{3}$ phonon. Additionally, this feature does not show any dependence on the applied magnetic field (see Supplementary materials Fig.~S4). We cannot conclude with the present data on the origin of this feature which appears as characteristic of the low temperature magnetic phase in bulk CrSBr.

\section{Conclusions}

In summary, using temperature and magnetic field dependent Raman scattering techniques, we have described the strong spin-phonon coupling in bulk CrSBr. The different magnetic phases have distinct Raman scattering signatures that experiments performed in a tilted geometry allow to precisely trace and investigate. Our ab-initio calculations of the phonon band structure in the AF phase agree well with the Raman scattering experiments and allow for the identification of the different phonon modes. We have shown the Raman scattering signatures of the still-debated magnetic phase at low temperature in the form of discontinuities in the temperature evolution of the energy and linewidth of the $A_{1g}^3$ phonon mode, in the appearance of a new Raman scattering peak at $E=467$~cm$^{-1}$. Raman scattering technique applied to magnetic layered materials are very efficient to get insights into the rich magnetic phases of system with magnetization aligned in the plane of the layers.

\begin{acknowledgements}

We acknowledge technical support from I. Breslavetz. This work has been partially supported by the EC Graphene Flagship project. N. P. W. acknowledges support from the Deutsche Forschungsgemeinschaft (DFG, German Research Foundation) under Germany’s Excellence Strategy—EXC-2111—390814868. Z.S. was supported by project LTAUSA19034 from Ministry of Education Youth and Sports (MEYS). R.H. acknowledges support by the state of Baden-W\"{u}rttemberg through bwHPC.

\end{acknowledgements}


%

\pagebreak

\widetext

\begin{center}
\textbf{\large Supplemental informations for Raman scattering signatures of the strong spin-phonon coupling in the bulk magnetic van der Waals material CrSBr}
\end{center}

\setcounter{equation}{0}
\setcounter{figure}{0}
\setcounter{table}{0}
\setcounter{page}{1}
\makeatletter
\renewcommand{\theequation}{S\arabic{equation}}
\renewcommand{\thefigure}{S\arabic{figure}}
\renewcommand{\thetable}{S\arabic{table}}
\renewcommand{\bibnumfmt}[1]{[S#1]}
\renewcommand{\citenumfont}[1]{S#1}

In this supplementary materials, we provide a table comparing our phonon calculations with experimental observations, the experimentally observed evolution with temperature of energy positions, linewidth, and integrated intensities, individual raman scattering spectra of the temperature dependence, and the magneto-Raman scattering response up to $B=1$~T including the second order spectrum.

\clearpage

\begingroup
\squeezetable
\begin{table}
\setlength{\tabcolsep}{20pt}
\renewcommand{\arraystretch}{1.5}

\caption {\label{tab:table1} Summary of experimental Raman peak position and calculated Raman frequencies and assigned vibration modes for CrSBr. \\}

\begin{ruledtabular}
\begin{tabular}{llll}
 Experimental & \multicolumn{2}{c}{Calculated} & Vibration mode \\
\cline{2-3}
        &  $\Gamma$ point   & Z point  \\    \hline
 $57.42  \pm 0.03$   &  ---   & 56.5  &  $B^{Z}_{3g}$    \\

 $71.35  \pm 0.09$    &  73.6   & ---      &  $B^{1}_{3g}$    \\

 $95.04  \pm 0.01$    &  90.8   & 83.9     &  $B^{1}_{2g}$    \\

 $121.37 \pm 0.02$   &  120.0   & 109.3   &  $A^{1}_{1g}$    \\

 $194.70 \pm 0.13$    &  195.2   & 191.5   &  $B^{2}_{3g}$    \\

 $199.20  \pm 0.04$   &  193.2   & 193.7   &  $B^{2}_{2g}$    \\

 $254.25 \pm 0.01$   &  246.9   & 245.8   &  $A^{2}_{1g}$    \\

 $304.83 \pm 0.26$    &  302.4   & 302.3   &  $B^{3}_{2g}$    \\

 $346.18 \pm 0.01$    &  345.1   & 344.8   &  $A^{3}_{1g}$    \\

 $367.50 \pm 0.03$    &  385.1   & 385.0   &  $B^{3}_{3g}$    \\

\end{tabular}
\end{ruledtabular}
\end{table}
\endgroup

\clearpage

\begin{figure}
\includegraphics[width=1\linewidth,angle=0,clip]{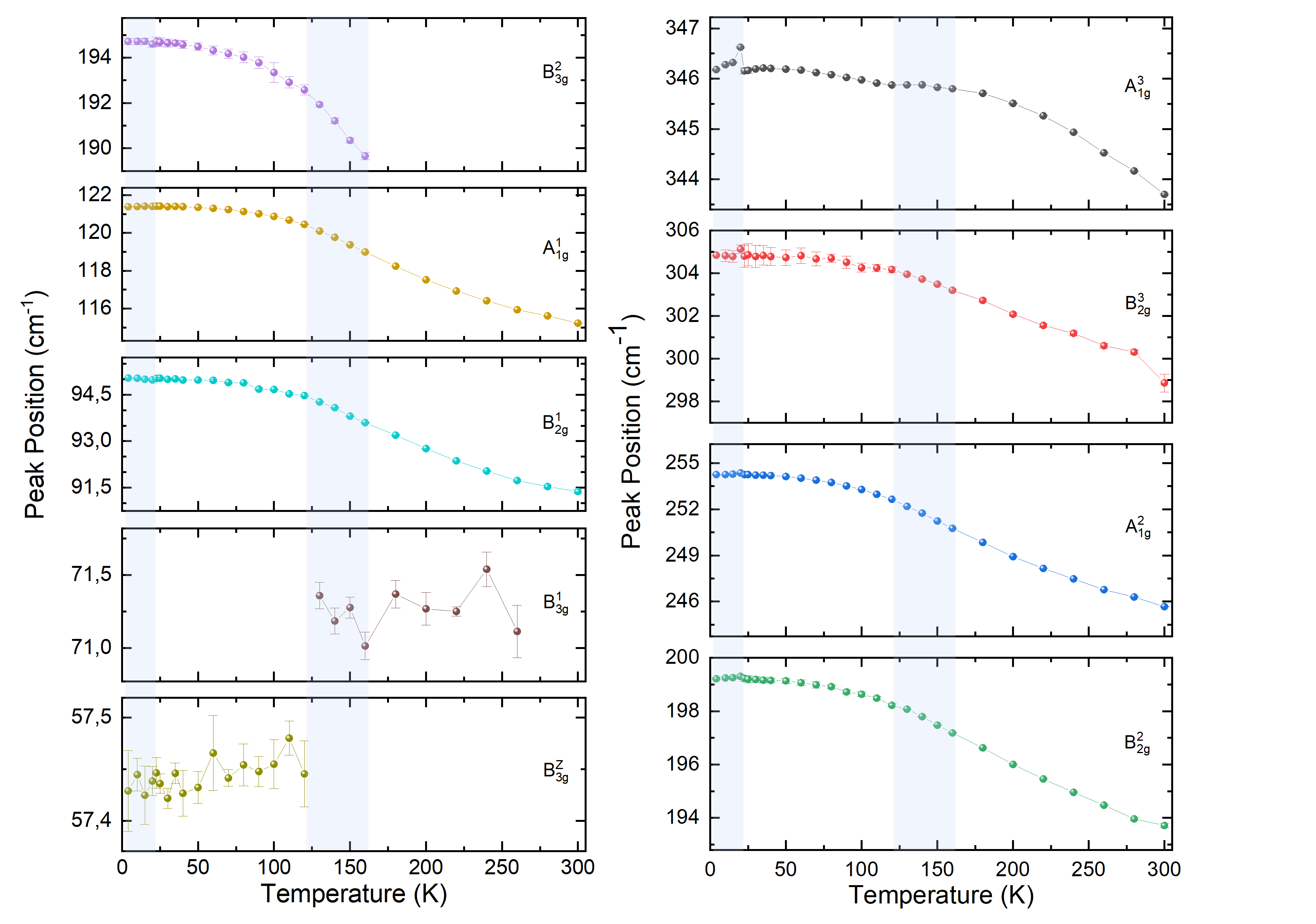}
\caption{Evolution of the energy position of all observed Raman scattering peaks of bulk CrSBr as a function of temperature.
\label{FigS1}}
\end{figure}

\clearpage

\begin{figure}
\includegraphics[width=1\linewidth,angle=0,clip]{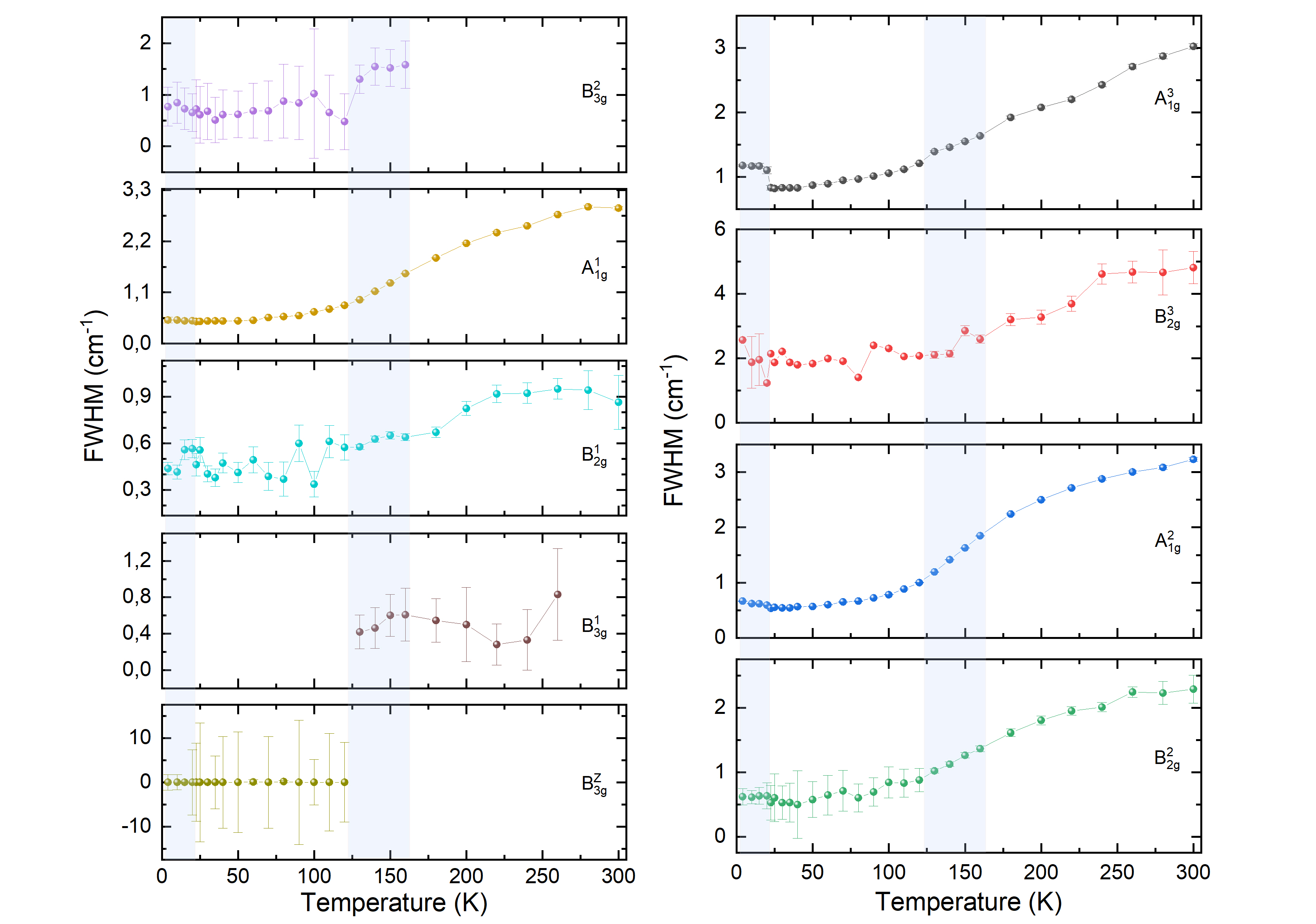}
\caption{Evolution of the full width at half maximum of all observed Raman scattering peaks of bulk CrSBr as a function of temperature.
\label{FigS2}}
\end{figure}

\clearpage

\begin{figure}
\includegraphics[width=1\linewidth,angle=0,clip]{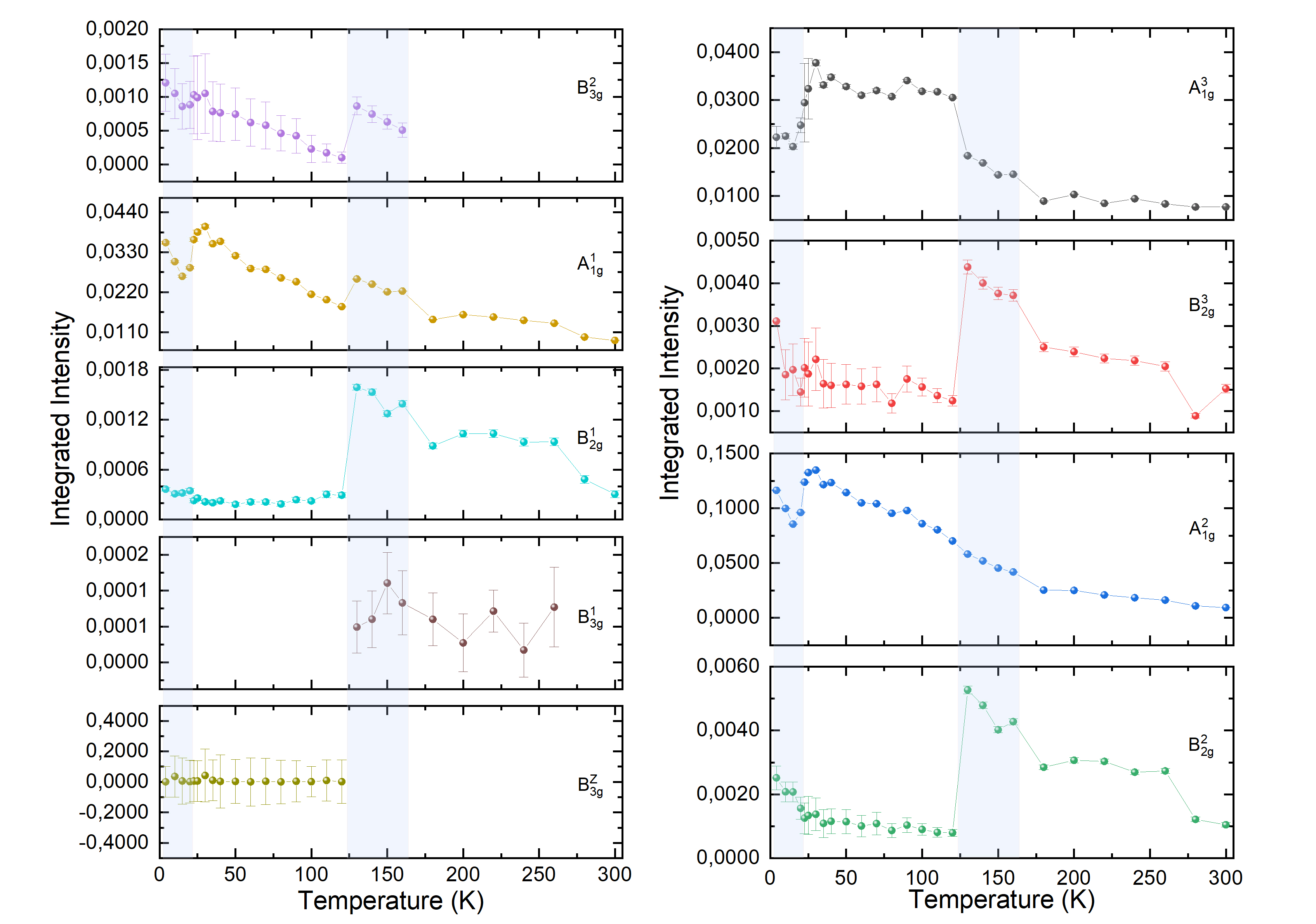}
\caption{Evolution of the integrated intensity of all observed Raman scattering peaks of bulk CrSBr as a function of temperature.
\label{FigS2}}
\end{figure}

\clearpage

\begin{figure}
\includegraphics[width=1\linewidth,angle=0,clip]{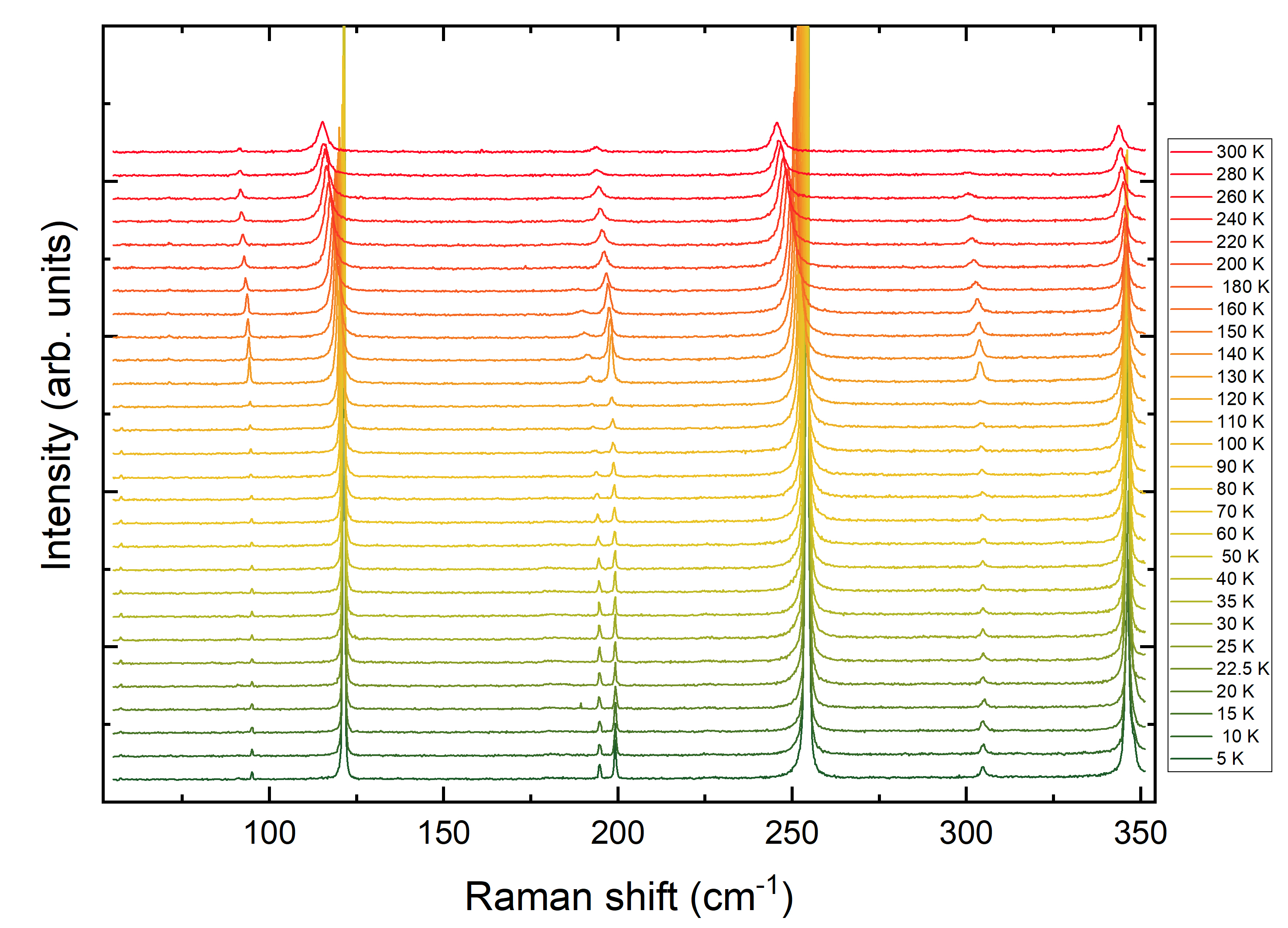}
\caption{Unpolarized Raman scattering response for different temperatures from room temperature to liquid helium temperature.
\label{FigS2}}
\end{figure}

\clearpage

\begin{figure}
\includegraphics[width=1\linewidth,angle=0,clip]{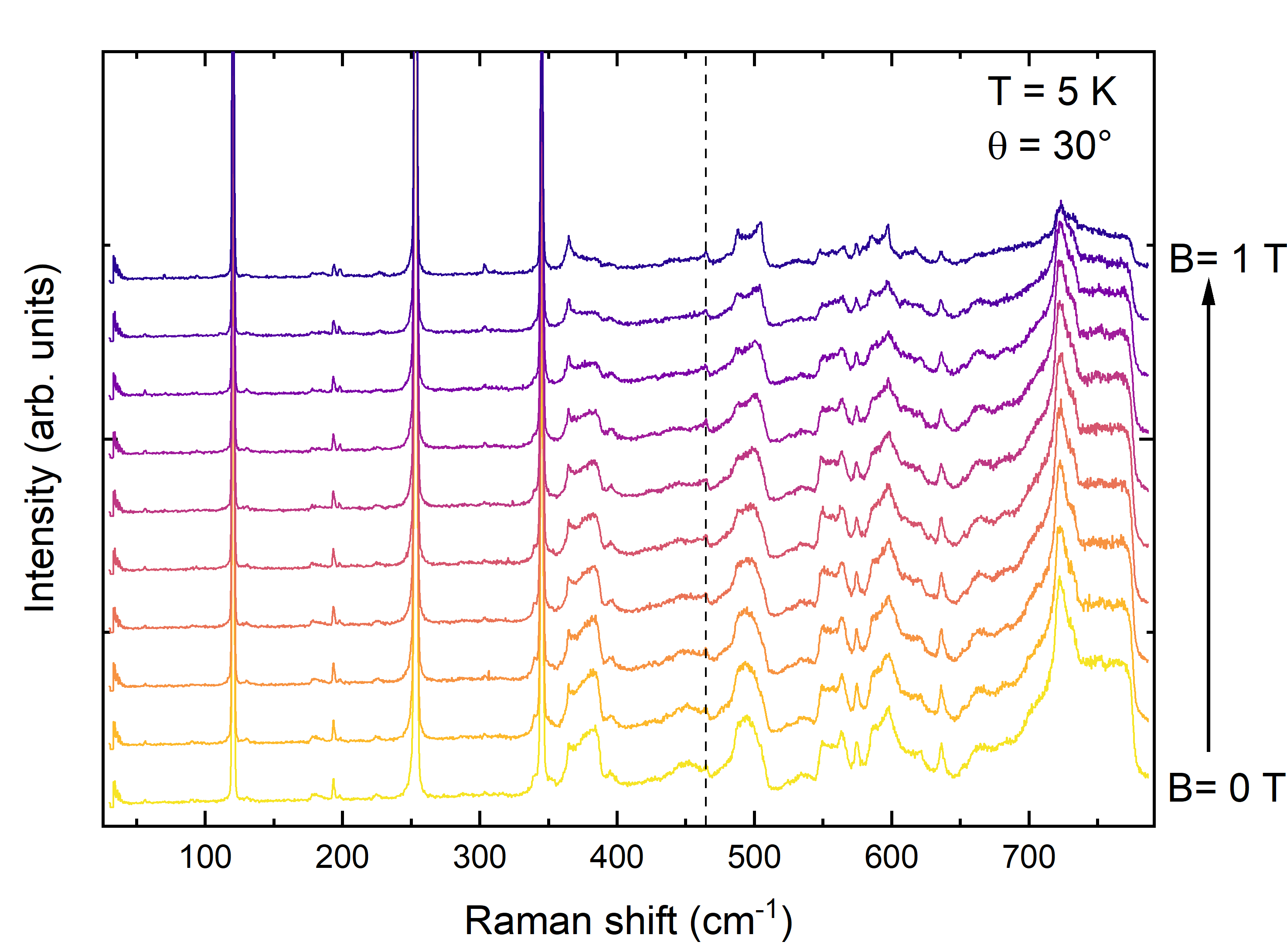}
\caption{Low temperature ($T=5$~K) unpolarized Raman scattering response of bulk CrSBr for selected values of the magnetic field up to $B=1$~T. The sample c axis is tilted by $30^{\circ}$ with respect to the external magnetic field and to the light propagation direction. The vertical dashed line indicates the $467$~cm$^{-1}$ Raman scattering peak that appears in low temperature magnetic phase. The second order Raman scattering spectrum evolves with the applied magnetic field
\label{FigSBfiled}}
\end{figure}

\end{document}